\documentclass[10pt,conference]{IEEEtran}

\usepackage{hyperref}
\usepackage{paralist}
\usepackage{color} 
\usepackage{verbatim}

\usepackage{listings}

\title{Coming: a Tool for Mining Change Pattern Instances from Git Commits}

\author{\IEEEauthorblockN{ Matias Martinez}
\IEEEauthorblockA{{University of Valenciennes, France}}\\
\and
\IEEEauthorblockN{ Martin Monperrus}
\IEEEauthorblockA{{KTH, Sweden} }
}

\usepackage{url}

\begin{document}

\maketitle

\begin{abstract}

Software repositories such as Git have become a relevant source of information for software engineer researcher. 
For instance, the detection of Commits that fulfill a given criterion (e.g., bugfixing commits) is one of the most frequent tasks done to understand the software evolution.
However, to our knowledge, there is not open-source tools that, given a Git repository, 
returns all the instances of a given change pattern. 
 
In this paper we present Coming, a tool that takes an input a Git repository and mines instances of change patterns on each commit.
For that, Coming computes fine-grained changes between two consecutive revisions, analyzes those changes to detect if they correspond to an instance of a change pattern (specified by the user using XML), 
and finally, after analyzing all the commits, it presents 
\begin{inparaenum}[\it a)]
\item the frequency of code changes and 
\item the instances found on each commit.
\end{inparaenum}

We evaluate Coming on a set of 28 pairs of revisions from Defects4J, finding instances of change patterns that involve \emph{If} conditions on 26 of them.

\end{abstract}

\section{Introduction}

During recent years software engineering researchers have been inspecting software code repositories such as Git or SVN to gain knowledge about the evolution of applications.
For example, there are a considerable number of studies (\cite{Hanam:2016:DBP, Pan:2009:TUB,Zhong:2015:ESR}) that have focused on the bug fixing activity by studying bug fix commits.
Other researcher have also presented approaches that aim at repairing automatically buggy programs.
Some of those repair approaches \cite{Long:2016:APG, le2016hdrepair} consume information  extracted from software repositories such as the most frequent bug fix patterns \cite{Kim:2013:APG} or frequency of code changes \cite{Soto:2018:CSK}.

To carry out such kind of analysis on repositories, a researcher needs a tool that:
\begin{inparaenum}[\it a)]
\item visits a set of revisions (i.e., commits);
\item filters those revisions that are interesting (e.g., bug fix commits) according to, for example, the commit message;
\item computes changes between a revision and its precedent;
\item summarizes the results (e.g., to compute the probability of each change type);
\item capture the commits that introduces a set of particular changes; 
among other activities.
\end{inparaenum}
However, to our knowledge, there is not an open-source that carries out all of these tasks.

In this paper we present \emph{Coming}, a tool that inspects Git repositories with two main goals:
\begin{inparaenum}[\it 1)]
\item to compute fine-grained changes between revisions, and 
\item to detect instances of change patter.
\end{inparaenum}

In a nutshell, 
Coming takes as inputs a list of revisions (e.g., Commits from git).
For each pair of consecutive revisions, i.e., $r$ and $r+1$, 
Coming first computes the changes between them at a fine-grained level using an AST-diff algorithm.
Then, it analyzes the changes to detect if they correspond to a change pattern given as input.
A change pattern specifies a set of changes (e.g., insert, remove) done over code entities (e.g., invocations, assignments).
A pair of revisions  has an instance of a pattern if:  
\begin{inparaenum}[\it 1)]
\item all the changes that a pattern specifies exist on the diff between those revisions;
\item the entities affected by the changes from the diff match with those that the pattern specifies.
\end{inparaenum}
Finally, after analyzing all the commits, 
Coming post-processes the results from each pair of revisions and exports the final results (i.e., pattern instances found and frequency of code changes) to a JSON format.
Moreover, Coming provides extension points for overriding the default behaviour or to add new functionality.

Coming can be used by researchers that aim at filtering commits to automatically create, for instance,  datasets of bugs. 
Moreover, it can be used by researchers that aim at post-processing a distilled set of changes found by the tool, and  then apply, for instance, an algorithm of change clustering.

To evaluate Coming,  we first collect 28 pairs of revisions from Defects4J \cite{Just:2014:DDE} which diffs affected to, at least, one \emph{if} condition.
Then, we write change patterns that specify different changes over \emph{if}. 
Finally, we execute Coming to detect instances of such patterns over the 28 pairs of revisions.
Coming could successfully find the correct instance of a change pattern on 26 pairs.

Coming is publicly available at \url{https://github.com/Spirals-Team/coming}.
The video that shows a demonstration of Coming is available at \url{https://youtu.be/dR6B9qRpjic}

\section{Approach}

\subsection{Goals of Coming}

The main goals of Coming are:
\begin{inparaenum}[\it a)]
\item to compute the fine-grain changes between two revisions;
\item to detect instances of change pattern from those fine-grain changes;
\item to count the frequency of changes along all the revisions of a Git repository; and
\item to count the occurrence of change pattern instances.
\end{inparaenum}

In the rest of this section, we present the series of steps that Coming carries out to accomplish those goals.

\subsection{Coming Inputs}
\label{sec:approach_input}
\label{sec:input:git}

Coming analyzes commits from a Git repository, whose path is given as parameter.
The implementation of Coming navigates the Git history using the library Eclipse GIT.\footnote{\url{https://projects.eclipse.org/projects/technology.egit}}
Coming navigates each commit starting from the oldest one.
For each commit $c$, Coming takes each  file $f$  that the commit  $c$ modifies, and creates a revision pair with:
\begin{inparaenum}[\it a)]
\item a file $f$ modified by $C$, and 
\item the previous version of $f$, introduced or modified by a commit older than $C$.
\end{inparaenum}

\subsection{AST-based Analysis of Revisions}
\label{sec:comparison:ast}

Coming carries out a fine-grained comparisons based on the AST (Abstract Syntax Tree) of each revision pair. 
This step has two main steps:

\subsubsection{AST Representation of Files}
\label{sec:modelization}

Coming creates a AST from the code of a revision.
In a AST, tree node corresponds to a code element (e.g., an invocation, a parameter).
Previous works have been working on different granularities of tree nodes: 
from coarse-grained from ChangeDistiller \cite{Gall:2009:CAE}, where the finest-grained code element represented by a node is the statement, 
to fine-grained such as Eclipse JDT \cite{Falleri:2014:FAS}.

In this paper,
we present a new level of granularity of AST, named GTSpoon, which is based on the Spoon meta-model.\footnote{Spoon meta-model: \url{http://spoon.gforge.inria.fr/code_elements.html} and \url{http://spoon.gforge.inria.fr/structural_elements.html}}
Spoon \cite{Pawlak:2016:SLI} is a library to analyze, transform, rewrite, transpile Java source code.

The granularity-level of the Spoon meta-model is between the previously two mentioned:
\begin{inparaenum}[\it a)]
\item Spoon nodes are finer (e.g., parameter, field write and read) that ChangeDistiller (statements); and 
\item those nodes contain more information than the Eclipse JDT nodes,  resulting  more compact trees.
\footnote{Discussion about the Spoon and JDT granularity:  \url{https://github.com/INRIA/spoon/issues/1303}}
\end{inparaenum}

 Coming uses, by default, the GTSpoon granularity, which means that each node of the AST of a revision $r$ is corresponds to an element from the Spoon model of $r$.
 We have written an open-source library named GTSpoon\footnote{\url{https://github.com/SpoonLabs/gumtree-spoon-ast-diff}} that returns a AST with that granularity from a source code file.

 \begin{comment}
 In our opinion, the advantage to use this level of granularity is two fold.
\begin{inparaenum}[\it 1)]
\item symbiotic with the automatic software approach such as Astor and Nopol, Genesis \cite{}.
That is, empirical results on, for instance, bug fix patterns can directly consumed by a repair approach.
\item AST compressible by humans.
\end{inparaenum}
 \end{comment}

\subsubsection{Tree-Diff Comparison}
\label{sec:comparison:tree}

To obtain the different  between two models $m_s$ and $m_t$ (by default GTSpoon's ASTs) retrieved from two revisions $r_s$ and $r_s$, resp., Coming applies a tree-difference algorithm. 
By default, Coming uses GumTree \cite{Falleri:2014:FAS}, a state-of-the-art AST diff algorithm.
The output of the diff between $m_s$ and $m_t$  is a list of \emph{Operations}, where each of them contains 
\begin{inparaenum}[\it a)]
\item an action type (Insert, Remove, Update, Move);
\item a reference to a node from $m_s$; and/or
\item a reference to a node from $m_t$.
\end{inparaenum}

\subsection{Analysis of Diffs: Finding Instances of Change Pattern}

Coming executes a set of \emph{Analyzers} which take as input the results of  previous diffs and carry out some task.
In this paper we present an analyzer that mines instances of change patterns.
The analyzer uses (and improves) the specification of change pattern that we have  previously defined \cite{Martinez:2013:AEI}  and implements the  algorithm, also defined tehre,  to match patterns and changes.

A \emph{Change Pattern} defines a set of changes between two revisions and  the elements affected by those changes. 

A pattern has a list called \emph{Pattern actions} $pa$  where each of them specifies a particular change  between two revisions. 
A \emph{Pattern action} has two fields.
First, the type of action, with four predefined values: insert, mode, remove, and update.
Secondly, it contains a reference to a $Pattern Entity$ which has  also three fields: 
\begin{inparaenum}[\it 1)]
\item \emph{type}, which indicates the type of code element of the entity  (e.g., if, invocation,  return);
\item \emph{value}, which indicates the value of the element (e.g, $callMethod1()$, $return \quad null$);
\item \emph{parent},  a recursive relation to a $Pattern Entity$ which indicates the parent of an entity. This relation has an argument, \emph{distance}, which indicates the max distance between $e$ and $ep$ in the AST. For example, a value of 1 indicates that the $ep$ is the immediate parent of $e$, whereas 2 means a grand-parent relation. 
\end{inparaenum}
The use of  wildcard character ``*'' in any of those mentioned fields produces a matching with any kind of type or value.

Coming accepts Change Patterns specified in a XML files.
Listing \ref{lis:pattern1} shows as example a pattern in XML that specifies:
\begin{inparaenum}[\it a)]
\item two entities (id 1 and 2), one representing a $Return$, the second one an $If$; \item a parent relation between the $if$ and the  $Return$ entities (with a max distance of 2 nodes); and
\item two actions of type INS (insert), one affecting the entity id 1 ($Return$), the other one the entity id 2 (the $if$).
\end{inparaenum}

\lstset{language=XML}
\lstset{
    basicstyle=\scriptsize,
  columns=fullflexible,
  showstringspaces=false,
  commentstyle=\color{gray}\upshape
}
\begin{lstlisting}[caption=Change Pattern \emph{Add If-Return}, label=lis:pattern1]
<pattern>
	<entity id=``1" type=``Return">
		<parent parentId=``2" distance=``2" />
	</entity>
	<entity id=``2" type=``If" />
	<action entityId=``1" type=``INS" />
	<action entityId=``2" type=``INS" />
</pattern>
\end{lstlisting}

\vspace{-0.5cm}
\subsection{Summarization of Results}

Finally, Coming processes all results obtained from the analyzers over all the commits  and then it exports the results to a JSON file.

\subsection{Extending Coming}

Beyond the functionality that Coming already includes, such as AST differencing and mining of change pattern instances, 
it provides \emph{extension points} to override the default behaviour and to define new tools for evolution analysis. 
The main extension points that Coming provides are the following, with, in parentheses the implementation already provided.
\begin{inparaenum}[\it a)]
\item \emph{Input}  (Git, Files System); 
\item \emph{Revision filter}  (presence of keywords in revisions messages, size of the revisions in terms of \# of hunks and in terms of \# files);
\item\emph{Analyzers} (Computation of syntactical (line-based) diff, AST-based diff, pattern instance detection);
\item \emph{Output processor} (Standard output, JSON containing the  instances found and change frequency).
\end{inparaenum}

In the documentation hosted in the Coming Github site,\footnote{\url{https://github.com/Spirals-Team/coming/blob/master/docs/extension_points.md}}  we explain how to create an new implementation for each extension point.

\section{Evaluation}

This experiment aims at measuring the ability of Coming to detect change pattern instances.
For this propose,  we create a set of 10 patterns from the related work. Then we run Coming over 28 pairs of revisions to mine instances of those patterns.

\subsection{Experiment setup}

In this experiment, we aim at mining instances of change pattern detecting bug fixing.
We consider Defects4J \cite{Just:2014:DDE}, a dataset of buggy programs from 6 Java open-source projects.
It
 contains, for each buggy program, a patch that repairs the bug. 
Due to the scope of this paper, we focus on buggy programs:
\begin{inparaenum}[\it 1)]
\item from the Apache Commons Math project; and
\item whose patches affect \emph{if} conditions.
\end{inparaenum}
In total, with the help of the Defects4J dissection \cite{defects4J-dissection}, the number of buggy programs satisfying  those criteria is 28.
For each of those bugs, we prepare the buggy and the patched version according to Coming's input format.

Then, we create a set of 10 change patterns to detect the changes that affect the \emph{if} conditions.
For instance, the first pattern \emph{Add If-return} corresponds to that one presented in Listing \ref{lis:pattern1}. It is able to detect, for instance, the changes between the buggy and patched version of bug M3 from Defects4J, which patch is shown in Listing \ref{lis:patchM3}. 

\lstset{language=Java}
\lstset{
    basicstyle=\scriptsize,
  columns=fullflexible,
  commentstyle=\color{gray}\upshape
}
\begin{lstlisting}[caption=Bug fix changes corresponding to bug Math-3. It is an instance of  Change Pattern \emph{Add If-Return} presented in Listing \ref{lis:pattern1}., label=lis:patchM3]

@@ -818,10 +818,7 @@ public class MathArrays {
+       if (len == 1) {
+            return a[0] * b[0];
+        }

\end{lstlisting}

Finally, we execute Coming over the 28 pairs of buggy and patched revisions from Commons Math projects.
We then manually inspect the results i.e., the mined instances from the revisions, to assert whether they are:
\begin{inparaenum}[\it a)]
\item \emph{true positive} i.e., the pattern instance exists between the revisions,
\item \emph{false negatives} i.e., Coming could not detect an instance of the pattern.
\end{inparaenum}

\subsection{Experimental Results}
\begin{table}[t!]
\caption{Mining Change pattern instances over buggy and patched version of Defects4J defects. The columns TP, FN and TN  shows the true positives, false negatives and true negatives.
A Defects4J's revision could have +1 pattern instances.}
\centering
 \begin{tabular}{|c |c | c| c|} 
 \hline
Change Pattern & TP (instances found) & FN &TN  \\ 
 \hline
 \hline
Add If-return&M3, M38, M53,M55,&M60&-\\
& M84, M92, M93 &&\\
\hline

Add If-return null&M4&-&-\\
 \hline
Add If-assig&M29, M51, M54, M102&-&-\\
 \hline
Add If-throw&M19, M25, M45, M48, M73, M99&-&M86\\
 \hline
Upd If-cond&M21, M37&-&-\\
 \hline
Add 2 nested Ifs&M39, M68, M78&M64&-\\
  \hline
Mov If-return&M64&-&-\\
 \hline

Add If-break&M1&-&-\\
  \hline
Del If-return&-&-&M64\\
 \hline
Add If Mov assig&M95&-&-\\

  \hline
 \end{tabular}
 \label{tab:instances}
\end{table}
 
Table \ref{tab:instances} shows the results of our experiment. 
The first column shows the change pattern name.

The second column (TP) shows, for each pattern $p$, 
the Defects4J \emph{identifier} for which Coming can successfully find a pattern instance of $p$ between the buggy and the patched version.
For example, Coming correctly identifies an instance of pattern ``Add-If-Return" in revision Math-3. 
In total, Coming can find correctly instances for 26 out of 28 patches (93\%), those are true positives. 
Moreover, Coming is  capable of finding more than 1 instances of the same pattern inside a revision pair. For instance, the revisions Math-93 has two instances of pattern ``If-Return''.

Then, the column FN shows the false negatives, i.e., the revisions that actually have an pattern instance but Coming fails to detect it.
We observe that the two false negatives are due the AST diff algorithm (Gumtree in vanilla mode) which did not create a correct minimal diff, i.e., it produces unnecessary INS and DELETE operations.

Finally, the column TN shows the cases considered as true negative, i.e., Coming does not return any instance (correctly).
The line line-based diff shows that \emph{if} conditions are added and removed, but Coming does not detect any instance of patterns \emph{Add If-*} or \emph{Del if-*}
A true negative occurs in Math-64, which patch is partially presented in Listing \ref{lis:m64}.

\begin{lstlisting}[caption=Two hunks from the patch Math-64. The Tree diff algorithm detects that the if condition is moved, label=lis:m64]

+                    	if (checker.converged(getIterations(), previous, current)) {
+                    		return current;
+                    	}
+                    }
-               } else {
-                      if (checker.converged(getIterations(), previous, current)) {
-                        return current;
-                    }
\end{lstlisting}

The listing shows two hunks, one that adds an \emph{if}, another that removes the same \emph{if} code.
From that revision pair, the AST-diff algorithm Gumtree detects  \emph{move} operations (both the \emph{If} and \emph{return} elements  are moved to another location).
Thus,  Coming is not able to find an instance of the pattern \emph{Del If-return} giving those two AST changes (Moves).
Consistently, when Coming mines instances of the pattern \emph{Mov If-return}, it successfully finds one between the buggy and patched version of Math-64.

Lastly, Table \ref{tab:instances} shows a pattern ``Add If-return null'' that specify the value of the entity, in addition to the entity type.
Using this feature, Coming can identify, for instance in Math-4, an instance of an \emph{if} that returns a $null$ value.

The code base of Coming includes the specifications of all the patterns presented in this experiment.\footnote{\url{https://github.com/Spirals-Team/coming/blob/master/docs/experiment_mining_instances_d4j.md}}

\section{Related work}

Coming  uses the method to specify a change pattern  that we presented in \cite{Martinez:2013:AEI} and implements the instance mining algorithm presented in that work.
Moreover, Coming provides several improvements including:
\begin{inparaenum}[\it 1)]
\item a finer-grained level of ASTs, which allows to create more precise change pattern;
\item matching of entity values;
\item more descriptive parent relation (allowing a chain of parents);
\item the use of a more reliable tree-diff algorithm \cite{Falleri:2014:FAS}.
\end{inparaenum}

There are other open-source tools that focus on the analysis of software repositories such as Gits.
Some of them  are: PyDriller\footnote{https://github.com/ishepard/pydriller} \cite{PyDriller},
Git-of-theseus\footnote{https://github.com/erikbern/git-of-theseus/},
CVSAnalY\footnote{https://github.com/MetricsGrimoire/CVSAnalY} and 
Hercules\footnote{https://github.com/src-d/hercules}.
However, to our knowledge, these tools do neither provide a fine-grained analysis of changes between revisions, nor the detection of change instances.

Different approaches have focused on the mining of bug fix pattern.
For instance, Madeiral et al. \cite{madeiral:hal-01851813} have presented an approach that detects repair patterns in patches, which performs source code change analysis at abstract-syntax tree level.
Their approach, as it is also built over our technology stack (GTSpoon, Spoon and GumTree), could be easily included in Coming using the extension point \emph{output  processor}.
Osman et al. \cite{Osman2014} analyze code hunks from line-based diff to detect bug-fix patterns,  Rolim et al. \cite{Rolim2018} propose a method for discovering quick fixes based on the identification of code edits (at the level of AST) from revisions, and then to cluster those edits. A similar work has been done by Molderez et al. \cite{Molderez2017} which use closed frequent itemset mining algorithm on sets of distilled  code changes.
Hanam et al. \cite{Hanam:2016:DBP} present a tool for discovering the most prevalent and detectable bug patterns on JavaScript code, based on unsupervised machine learning. 
As difference of those works, Coming focuses on the detection of \emph{instances} of existing change pattern, including bug fix pattern.
Moreover, as Coming computes the fine-grained diffs and also provides extension points to analyze that information, any of those works can be implemented in our tool.

\section{Future Work}
The current version of Coming includes all features presented in this paper.
Nevertheless, we continue working on new features and improving the tool usability.
Some of the planned features are:
\begin{inparaenum}[\it a)]
\item Enrichment of the pattern specification to include cardinality of elements (numbers of children, siblings, etc), assert the absence of elements, different matching strategies of entity types and values, and accept changes that affect different files;
\item parallelisation;
\item post-processors to mine, for instance, change patterns;
\item tuning arguments of tree-diff algorithm to avoid true negatives.
\end{inparaenum}
 
\section{Conclusion}
\label{sec:conclusion}
In this paper we present the tool named Coming which given a Git repository, navigates every commit, calculates fine-grained changes between a revising of a commit and its precedent, detects change pattern instances from those changes, computes the frequency of code changes along the repository and finally exports the results in JSON format.
Coming presents extension points allowing researchers to plug-in their own approaches that, for example, focus on the discovering of bug-fix patterns from the changes computed by Coming.
Coming is publicly available at \url{https://github.com/Spirals-Team/coming/}. New features and extensions are welcome via Pull Request. 

\bibliographystyle{plain}
\bibliography{references}

\end{document}